\pdfoutput=1
\documentclass[runningheads]{llncs}
\usepackage[T1]{fontenc}
\usepackage{graphicx}
\usepackage{xcolor}
\usepackage{booktabs}
\usepackage{tikz}
\usetikzlibrary{shapes, shapes.geometric, arrows, positioning}
\usepackage{amsmath}
\usepackage{tikz}
\usetikzlibrary{shapes, arrows, positioning}
\usepackage{amssymb}
\usepackage{etoolbox}
\makeatletter
\patchcmd{\@startsection}{3.25ex}{2ex}{}{}
\makeatother
\tikzset{
 data/.style={
        trapezium, trapezium left angle=70, trapezium right angle=110,
        draw, fill=orange!15, font=\footnotesize\bfseries,
        minimum height=1.8em, inner sep=3pt, text centered
    },
    step/.style={
        rectangle, draw, rounded corners=3pt,
        minimum height=2.8em, text width=7.5em, text centered,
        font=\footnotesize, fill=#1
    },
    output/.style={
        circle, draw, 
        fill=green!12, font=\footnotesize\bfseries,
        inner sep=4pt
    },
    arrow/.style={->, thick, color=gray!70},
    darrow/.style={->, thick, dashed, color=red!50},
    pathlabel/.style={font=\tiny\itshape, text=gray!80}
}

\begin{document}
\title{MCQ Difficulty Prediction via Modeling Learner Heterogeneity Using Data-Driven Cognitive Profiling}

\titlerunning{MCQ Difficulty Prediction through Modeling Learner Heterogeneity}
%
\author{Dhriti Krishnan\inst{1}\orcidID{0009-0008-6508-9460} \and
Jaromir Savelka\inst{1}\orcidID{0000-0002-3674-5456}}

\institute{
Carnegie Mellon University, Pittsburgh, PA, USA\\
\email{dhritik@andrew.cmu.edu}, \email{jsavelka@andrew.cmu.edu} 
}

\authorrunning{D. Krishnan and J. Savelka}
%
%
\maketitle              
\begin{abstract}
Predicting the difficulty of multiple-choice questions (MCQs) is important for effective assessment, yet current methods typically assume a unimodal student ability distribution, overlooking the heterogeneous nature of student misconceptions. We propose a persona-driven framework that replaces
theoretical ability sampling with data-driven cognitive profiling. Using student interactions from the EEDI dataset, we identify behavioral personas via latent class analysis (LCA), then condition a large language model (LLM) to simulate response distributions for each persona. These signals are aggregated with topic context and fed into a Ridge Regression model to predict the item response theory (IRT) difficulty parameter. With five-fold cross-validation, our method improves over a recent baseline (MSE: $0.367 \rightarrow 0.274$;
$R^2$: $0.525 \rightarrow 0.686$). The discovered personas are interpretable and offer insights into why items are difficult, with potential applications
to diagnostic assessment design.
\end{abstract}

\section{Introduction}

Accurate estimates of item difficulty are fundamental to effective testing, item calibration, and curriculum design~\cite{frances2014item}. Traditional approaches based on item response theory (IRT)~\cite{rasch1960probabilistic,de2013theory} require substantial pretesting data, meaning newly authored items, particularly multiple-choice questions (MCQs)~\cite{butler2018multiple}, cannot be deployed until sufficient student responses have been collected. Recent work has explored predicting difficulty without pretesting, by inferring it from the question text~\cite{huang2020difficulty,haetal2019predicting}, learned
semantic representations~\cite{scarlatos2023turning,feng2025reasoning}, or LLM-simulated student responses~\cite{heyueya2024psychometric,park2024llasa}. However, these approaches typically assume a unimodal student population, overlooking heterogeneity in learner behavior.

Our central hypothesis is that student errors are not random but reflect systematic patterns tied to distinct cognitive profiles. Learners with comparable overall performance may differ in the types of mistakes they make and the concepts they fail to apply~\cite{koedinger2004real}, and student
knowledge is better characterized by multiple knowledge components than a single ability parameter~\cite{koedinger2012kli}. If this heterogeneity is
ignored, simulated response distributions, and the difficulty estimates derived from them, may poorly reflect how diverse learner populations interact with assessment items.

Our contributions presented in this work are as follows:
\begin{enumerate}
    \item We propose a novel persona-driven framework that (i) uses Latent Class Analysis (LCA) on student response data to discover interpretable behavioral personas that capture how different types of learners systematically differ in their knowledge and misconceptions; (ii) conditions an LLM on these personas to simulate response distributions, replacing the standard assumption of a unimodal ability distribution.
    \item We demonstrate improved prediction of IRT difficulty parameters over a recent state-of-the-art baseline~\cite{feng2025reasoning} on the EEDI dataset using five-fold cross-validation (MSE: $0.471 \rightarrow 0.274$;
    $R^2$: $0.525 \rightarrow 0.686$), while producing interpretable personas with potential applications to diagnostic assessment design.
\end{enumerate}

\section{Related Work}
\label{sec:methods}

\paragraph{\textbf{LLMs as Simulated Students}}
He-Yueya et al.~\cite{heyueya2024psychometric} introduced psychometric
alignment as a metric for quantifying how well LLMs reflect human knowledge distributions. Park et al.~\cite{park2024llasa} demonstrated that LLMs can exhibit graded proficiency levels in a zero-shot setting without fine-tuning, suggesting their potential as proxies for learners at different ability levels. Building on this, Hu and Collier~\cite{hu2024quantifying} showed that conditioning LLMs on persona variables systematically alters response patterns, while Yuan et al.~\cite{yuan2026validstudentsimulationlarge} argued that valid student simulation requires explicitly specifying the epistemic scope of a simulated learner's knowledge. In our work, we extend this line of research by deriving behavioral personas directly from observed student response patterns to ensure that the simulated students reflect the actual capabilities and misconceptions found in real-world classrooms.

\vspace{-0.7\baselineskip}
\paragraph{\textbf{Modeling Learner Ability}}
LLM-based difficulty prediction has gained attention through the BEA 2024 Shared Task~\cite{yaneva2024findings}, where participating systems showed competitive zero-shot performance but generally struggled to model distractor plausibility. Subsequent work has focused on how to represent variation in student ability within simulation pipelines. Feng et al.~\cite{feng2025reasoning} sampled scalar knowledge levels from a standard normal distribution, while Scarlatos et al.~\cite{scarlatos2025smart} used direct preference optimization to align simulated students with IRT ability parameters. These approaches implicitly assume that student errors vary in frequency rather than in kind. Our framework relaxes this assumption by conditioning on discrete cognitive profiles that capture qualitative differences in error patterns across learner subpopulations.
\vspace{-0.7\baselineskip}
\paragraph{\textbf{The Homogeneity Gap}}
A recurring finding across literature is that LLMs tend to produce expert-biased responses, generating unrealistically accurate answers rather than reflecting the error patterns of typical
learners~\cite{heyueya2024psychometric}. Feng et al.~\cite{feng2024exploring} observed a similar pattern in distractor generation: LLMs produce mathematically valid distractors but failed to target the misconceptions that real students actually held. Taken together, these results point to a fundamental mismatch between how LLMs reason and how students err. Our work tackles this by grounding simulation in cognitive personas discovered from actual student response data.

\section{Methodology}
\label{sec:methodology}

\begin{figure}[t]
\centering
\resizebox{\textwidth}{!}{
\begin{tikzpicture}[
    node distance=0.6cm and 0.9cm,
    >=latex
]

\node[step=purple!12] (llm) {
    \textbf{Step 2}\\[1pt] LLM Simulation\\[-1pt]
    {\tiny\textit{Claude 3.7 Sonnet}}
};
\node[step=teal!12, right=1.2cm of llm] (feats) {
    \textbf{Step 3}\\[1pt] Feature Extraction\\[-1pt]
    {\tiny\textit{$K \times 4$ probs + stats}}
};
\node[step=red!10, right=1.2cm of feats] (ridge) {
    \textbf{Step 4}\\[1pt] Ridge Regression\\[-1pt]
    {\tiny\textit{predict $\beta_i$}}
};

\node[step=blue!12, above=1.5cm of llm] (lca) {
    \textbf{Step 1}\\[1pt] LCA Clustering\\[-1pt]
    {\tiny\textit{5 personas}}
};
\node[step=gray!10, above=1.5cm of ridge] (irt) {
    \textbf{Step 0}\\[1pt] 2PL-IRT\\[-1pt]
    {\tiny\textit{ground truth}}
};

\path (lca.north) -- (irt.north) coordinate[midway] (midtop);
\node[data, above=0.8cm of midtop] (eedi) {EEDI Dataset};

\node[output, right=1cm of ridge] (pred) {$\hat{\beta}$};

\draw[arrow] (eedi.south) -- ++(0,-0.4) coordinate (fork);
\draw[arrow] (fork) -| (lca.north);
\draw[arrow] (fork) -| (irt.north);

\draw[arrow] (lca.south) -- (llm.north) node[midway, left, pathlabel] {personas};

\draw[arrow] (llm.east) -- (feats.west) node[midway, above, pathlabel] {prob.\ dists};
\draw[arrow] (feats.east) -- (ridge.west) node[midway, above, pathlabel] {features};
\draw[arrow] (ridge.east) -- (pred.west);

\draw[darrow] (irt.south) -- (ridge.north) node[midway, right, font=\tiny\itshape, text=red!60] {ground truth $\beta$};

\end{tikzpicture}
}

\caption{\textbf{The Proposed Pipeline.} LCA discovers 5 learner personas from student response data (Step~1). An LLM simulates per-persona response distributions (Step~2), which are aggregated into features (Step~3) and used to predict IRT difficulty via Ridge Regression (Step~4). Ground truth is estimated independently via 2PL-IRT (Step~0).}
\label{fig:pipeline}
\end{figure}
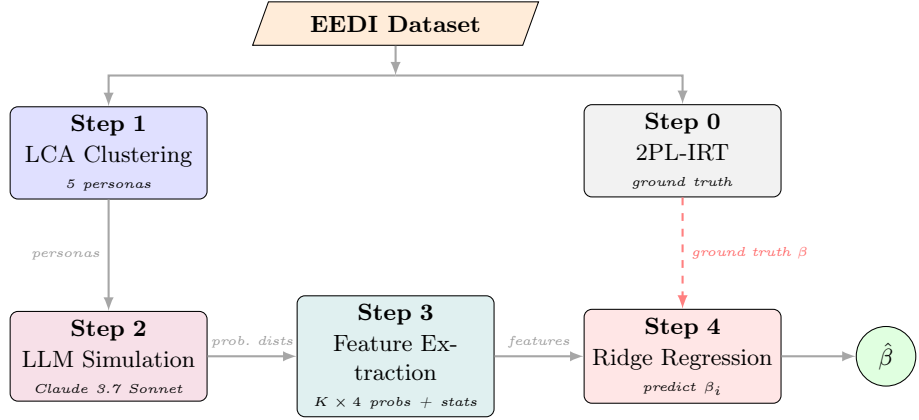

We propose a four-stage framework for estimating MCQ difficulty that
combines psychometric profiling with LLM-based student simulation:
(1)~data pre-processing and ground truth estimation,
(2)~psychometric profiling,
(3)~profile-conditioned response simulation,
(4)~predictive modeling of difficulty.
\subsection{Data Pre-processing and Ground Truth Estimation}

We use tasks 3 \& 4 of the EEDI Dataset~\cite{wang2020diagnostic}, which
contains 948 mathematics MCQs answered by 4,918 students. Each of the 1,382,728 interaction
records contains a student ID, question ID, selected answer option, and
binary correctness indicator. Questions are provided as images; we extract
their textual content using Tesseract OCR applied to grayscale-converted
images. Questions where OCR indicated
image-only content were excluded.

We partition the dataset into two disjoint subsets: a \textbf{profiling set} for discovering student personas, and an \textbf{estimation set} for ground truth estimation and difficulty prediction. For the profiling set, we apply dense core filtering, retaining questions with $\geq$50 responses and students with $\geq$10 attempts. For the estimation set, we retain questions with $\geq$20 responses, yielding 900 items. We fit a 2PL-IRT model on this subset using
py-irt~\cite{lalor2019emnlp},
estimating item difficulty ($\beta$) and discrimination ($\alpha$)
parameters that serve as ground truth targets.
\subsection{Psychometric Profiling}\label{sec:sec:psychometric}

\begin{figure}[t]
    \centering
    \includegraphics[width=0.995\textwidth]{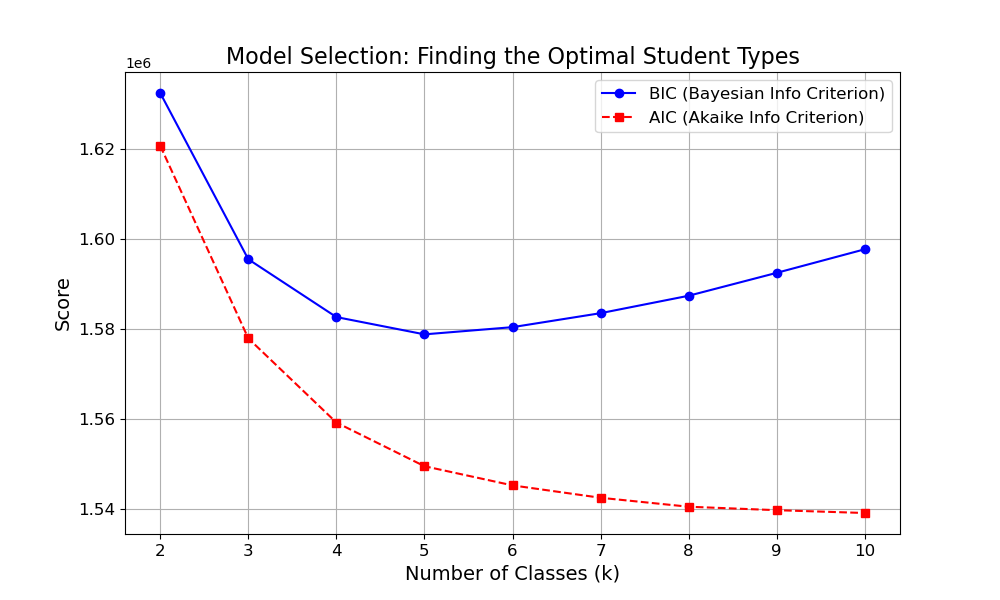}
    \caption{\textbf{Model selection for psychometric profiling.} Bayesian Information Criterion (BIC) and Akaike Information Criterion (AIC) for Latent Class models with varying numbers of latent classes. BIC reaches a global minimum at $k=5$, indicating the optimal number of learner profiles.}
    \label{fig:model_selection}
\end{figure}
We partition students into latent performance classes using LCA with a
binary measurement model (StepMix). Each student is assigned to a latent class $c_u \in \{1,\ldots,k\}$ based on maximum posterior
probability, and classes are ordered by mean accuracy.

To interpret each cluster, we compute a deviation score for each question
$i$ and cluster $c$:
\begin{equation}
\delta_i^{(c)} = a_i^{(c)} - \frac{1}{K}\sum_{k=1}^{K} a_i^{(k)},
\label{eq:deviation}
\end{equation}
where $a_i^{(c)}$ is the accuracy of cluster $c$ on question $i$. We
 select the 5 questions with the largest $\delta_i^{(c)}$ (strengths) and
the 5 with the most negative $\delta_i^{(c)}$ (weaknesses) per cluster.
These questions, with their topic labels and accuracy figures,
are provided to Claude Opus 4.5~\cite{anthropic2025opus45}, which produces
a persona name and description capturing the cognitive gap between each
cluster's strengths and weaknesses
(Table~\ref{tab:personas_summary}). For instance, ``The Conceptual
Reasoner'' can explain why dividing by a fraction increases a number but
cannot execute $2/9 \div 3/4$.

\subsection{Profile-Conditioned Response Simulation}\label{sec:sec:profile}

\begin{table}[t]
\renewcommand{\arraystretch}{1.5}
\centering
\caption{\textbf{Summary of Discovered Personas.} Five profiles with distinct procedural and conceptual capabilities.}
\label{tab:personas_summary}
\begin{tabular}{p{0.32\textwidth} p{0.32\textwidth} p{0.32\textwidth}}
\toprule
\textbf{Persona} & \textbf{Core Strength} & \textbf{Core Weakness} \\
\midrule
\textbf{The Rule Memorizer} & Applies formulas (e.g., multiplication) accurately. & Lacks number sense; misconceptions on magnitude. \\
\textbf{The Procedural\newline Calculator} & Strong on single-step arithmetic. & Fails inverse reasoning and multi-step logic. \\
\textbf{The Abstract Reasoner} & High logical/proportional reasoning. & Weak automaticity with basic arithmetic. \\
\textbf{The Conceptual\newline Reasoner} & Strong mathematical intuition (``Why''). & Breakdown in procedural fluency (``How''). \\
\textbf{The Fraction Calculator} & Solves standard fraction equations well. & Fails proportional reasoning in real contexts. \\
\bottomrule
\end{tabular}
\end{table}

Given an assessment item $Q_i$ and a persona description $\pi_c$, we need
to estimate how each type of learner would respond. A direct regression from
item text to difficulty would bypass the learner entirely; instead, we use
an LLM as a simulation engine because it can process both the question
content (provided as an image) and the persona description jointly,
producing response distributions that reflect persona-specific reasoning
patterns rather than optimal problem-solving.

Each question image is sent to Claude 3.7
Sonnet~\cite{anthropic2025sonnet37} along with a system prompt containing the
persona name and description. The model is instructed to estimate, as that
student type, the probability of selecting each answer option rather than
solve the item optimally. We denote the profile-conditioned option selection
distribution as:

\begin{equation}
\mathbf{p}_i^{(c)} =
\bigl[
p_{iA}^{(c)},\,
p_{iB}^{(c)},\,
p_{iC}^{(c)},\,
p_{iD}^{(c)}
\bigr],
\quad
\sum_{o \in \{A,B,C,D\}} p_{io}^{(c)} = 1.
\label{eq:profile_probs}
\end{equation}

This produces a $K \times 4$ matrix of option probabilities per item, where
$K{=}5$ personas each contribute a four-option distribution.
\subsection{Predictive Modeling of Item Difficulty}

For each item $i$, we construct a feature vector from the simulated
probabilities in Eq.~\ref{eq:profile_probs}. For each persona $c$, we
extract the probability assigned to the correct option,
$p_{i,\text{correct}}^{(c)}$, and compute aggregate statistics across
personas: mean, variance, and range of $p_{i,\text{correct}}^{(c)}$. Each
item's mathematical topic (Number, Algebra, or Geometry and Measure) is
one-hot encoded. All numeric features are standardised before modeling.

We predict the IRT difficulty parameter $\beta_i$ using Ridge
Regression~\cite{hoerl1970ridge} with cross-validated regularisation
strength ($\alpha \in \{0.1, 1, 10, 100, 500\}$).
\vspace{-0.7\baselineskip}

\section{Experiments}
\label{sec:results}

\subsection{Baseline Methods}

We compare against four baselines from Feng et
al.~\cite{feng2025reasoning}, evaluated on the same EEDI dataset:
\begin{itemize}
    \item \textbf{LR:} Linear regression on nine handcrafted syntactic and
    mathematical features.
    \item \textbf{FT:} Longformer-base-4096~\cite{beltagy2020longformer}
    finetuned to predict difficulty from tokenised question text.
    \item \textbf{FTWR:} FT augmented with GPT-4o-generated~\cite{hurst2024gpt}
    reasoning steps for the correct answer and feedback messages for each
    distractor.
    \item \textbf{Two-Stage Likelihood:} The primary method
    in~\cite{feng2025reasoning}. It samples student knowledge levels from a
    standard normal distribution, predicts per-option selection likelihoods,
    and regresses difficulty from these likelihoods.
\end{itemize}
\subsection{Results}
\label{sec:quant_results}
We use five-fold cross-validation, reporting mean and standard deviation
across folds. We evaluate with MSE and $R^2$.
Table~\ref{tab:results_main} summarises difficulty prediction performance.
Our method reduces MSE by 25\% relative to the strongest baseline and
improves $R^2$ from 0.525 to 0.686, explaining substantially more variance
in IRT-estimated difficulty.
\begin{table}[h!]
\centering
\renewcommand{\arraystretch}{1.3}
\setlength{\tabcolsep}{12pt}
\caption{Item difficulty prediction performance. Baseline results are
reported from \cite{feng2025reasoning}}
\label{tab:results_main}
\begin{tabular}{lcc}
\toprule
Method & MSE & $R^2$\\
\midrule
\textbf{Our method} & \textbf{0.274 $\pm$ 0.022} & \textbf{0.686 $\pm$ 0.012}\\
Two-Stage Likelihood~\cite{feng2025reasoning} & 0.367 $\pm$ 0.082 & 0.525 $\pm$ 0.101 \\
FTWR~\cite{feng2025reasoning} & 0.471 $\pm$ 0.149 & 0.390 $\pm$ 0.181 \\
FT~\cite{feng2025reasoning} & 0.522 $\pm$ 0.079 & 0.329 $\pm$ 0.048 \\
LR~\cite{feng2025reasoning} & 0.688 $\pm$ 0.028 & 0.084 $\pm$ 0.059\\
\bottomrule
\end{tabular}
\end{table}
\vspace{-0.7\baselineskip}
\section{Discussion}
\label{sec:discussion}

The improvement from $R^2{=}.525$ to $.686$ is consistent with our central
hypothesis: student errors on mathematical items are structured, not random.
Different learner profiles make different kinds of mistakes, and modelling
this explicitly produces simulated response distributions that better match
empirically estimated difficulties.

Our results also suggest that item difficulty is not an intrinsic property
of question text alone but emerges from the interaction between an item and
the learner population. An item requiring multi-step symbolic manipulation
may be straightforward for \textit{Rule Memorizers} but challenging for
\textit{Conceptual Reasoners}, even when both groups have similar overall
ability under a unidimensional IRT model~\cite{rasch1960probabilistic}. The
persona-conditioned simulation makes these differences visible and feeds
them into difficulty prediction.

To illustrate the depth of the discovered profiles,
Figure~\ref{fig:persona_spotlight} shows the full LLM-generated analysis
for ``The Conceptual Reasoner'' (Cluster 3). This student understands
mathematical logic but fails at symbolic execution---they can reason about
why dividing by a fraction increases a number, but cannot carry out the
procedure. Conditioning the LLM on this behavioural signature allows it to
simulate realistic ``right logic, wrong answer'' errors.
\vspace{-0.7\baselineskip}

\begin{figure}[t]
    \centering
    \fbox{
    \begin{minipage}{0.95\textwidth}
        \small
        \textbf{Persona Spotlight: Cluster 3 -- ``The Conceptual Reasoner''}

        \vspace{0.2cm}
        \textit{\textbf{LLM Analysis:}} The Conceptual Reasoner has
        developed strong mathematical intuition and can think logically
        about number relationships, proportional reasoning, and abstract
        concepts. They excel when problems require understanding `why'
        mathematics works --- reasoning about factors, analyzing how
        operations affect quantities, or thinking through sharing scenarios.
        However, they have critical gaps in procedural skills, particularly
        fraction operations: finding equivalent fractions, converting
        between representations, and applying the reciprocal method for
        division. The cognitive gap is striking --- they can reason
        abstractly about what happens when you divide by a fraction, but
        cannot execute $(2/9 \div 3/4)$.
    \end{minipage}
    }
    \caption{\textbf{LLM-Generated Description for ``The Conceptual
    Reasoner''.} The pipeline uses these descriptions to prompt the
    simulation LLM.}
    \label{fig:persona_spotlight}
\end{figure}

\subsection{Implications for Teaching Practice}

\paragraph{Cold-Start Problem}
Our framework enables approximate difficulty estimation at the time of item
authoring, without requiring pretesting data. This supports earlier
integration of new items into adaptive systems, particularly in domains
where content must be frequently refreshed.
\vspace{-0.7\baselineskip}
\paragraph{Diagnostic Item Design}
Rather than a single difficulty score, our approach yields profile-specific
response patterns showing which learner types are likely to struggle. This
helps instructors assess whether an item measures its intended concept or
instead reflects a specific procedural or representational challenge.
\vspace{-0.7\baselineskip}
\paragraph{Targeted Instruction}
Linking personas to error patterns helps interpret \textit{why} students
struggle. If \textit{Rule Memorizers} frequently fail an item, the item
likely depends on conceptual understanding rather than procedural recall---suggesting that conceptual explanations would be more useful than additional
drill.

\subsection{Limitations}

The EEDI dataset is specific to UK mathematics education, so the discovered
personas (e.g., ``The Fraction Calculator'') would not transfer directly to
other domains or curricula without re-running the profiling stage.
Our model also treats persona assignments as static, whereas real students
shift between profiles as they learn. Finally, baseline results are taken
directly from~\cite{feng2025reasoning}, whose setup differs from ours: they
use 327 MCQs (excluding diagram items) with a 65/15/20 train/validation/test
split, versus our 900 MCQs with five-fold cross-validation, and IRT ground
truth was estimated independently in both studies. These differences should
be kept in mind when interpreting the comparison.

\section{Conclusions and Future Work}
\label{sec:conclusions}
We presented a persona-driven framework for MCQ difficulty prediction that
derives learner profiles via LCA and conditions LLM simulation on these
profiles. On the EEDI dataset, this improves IRT difficulty estimation over
existing methods~\cite{feng2025reasoning}, suggesting that accounting for
learner heterogeneity matters more than increasing model complexity alone.

Future work includes applying the framework to subjects beyond mathematics
to test whether the discovered profiles are domain-specific, incorporating
temporal models such as knowledge tracing to capture how students shift
between profiles, and using persona-conditioned simulations to design
distractors that target specific misconceptions.

\begin{credits}
\subsubsection{\ackname}

This research is funded in part by the Carnegie Mellon-Accenture Center of Excellence in AI-Enabled Workforce Training (ACE-AI). The content of the information does not necessarily reflect the position or the policy of the funder and no official endorsement should be inferred. Generative AI was used in writing this article to improve surface language features.
\end{credits}

{\small
\bibliographystyle{splncs04}
\bibliography{sn-bibliography}

@inproceedings{feng2025reasoning,
  title={Reasoning and Sampling-Augmented MCQ Difficulty Prediction via LLMs},
  author={Feng, Wanyong and Tran, Peter and Sireci, Stephen and Lan, Andrew S.},
  booktitle={Artificial Intelligence in Education (AIED 2025), LNCS},
  pages={31--45},
  publisher={Springer},
  year={2025}
}

@article{koedinger2012kli,
  title={The Knowledge-Learning-Instruction Framework: Bridging the Science-Practice Chasm},
  author={Koedinger, Kenneth R and Corbett, Albert T and Perfetti, Charles},
  journal={Cognitive Science},
  volume={36},
  number={5},
  pages={757--798},
  year={2012}
}

@inproceedings{huang2020difficulty,
  title={Question Difficulty Prediction for Multiple Choice Problems in Medical Exams},
  author={Huang, Zhenya and Qi, Ye and Shen, Chuan and Ding, Gan},
  booktitle={Proceedings of CIKM},
  year={2019}
}

@inproceedings{scarlatos2025smart,
  title={SMART: Simulated Students Aligned with Item Response Theory for Question Difficulty Prediction},
  author={Scarlatos, Alexander and Fernandez, Nigel and Ormerod, Christopher and Lottridge, Susan and Lan, Andrew},
  booktitle={Proceedings of EMNLP 2025},
  year={2025}
}

@article{heyueya2024psychometric,
  title={Psychometric Alignment: Capturing Human Knowledge Distributions via Language Models},
  author={He-Yueya, Joy and Ma, Wanjing Anya and Gandhi, Kanishk and Domingue, Benjamin W. and Brunskill, Emma and Goodman, Noah D.},
  journal={arXiv preprint arXiv:2407.15645},
  year={2024}
}

@inproceedings{park2024llasa,
  title={Large Language Models are Students at Various Levels: Zero-shot Question Difficulty Estimation},
  author={Park, Jae-Woo and Park, Seong-Jin and Won, Hyun-Sik and Kim, Kang-Min},
  booktitle={Findings of EMNLP 2024},
  pages={8157--8177},
  year={2024}
}

@inproceedings{feng2024exploring,
  title={Exploring Automated Distractor Generation for Math Multiple-choice Questions via Large Language Models},
  author={Feng, Wanyong and Lee, Jaewook and McNichols, Hunter and Scarlatos, Alexander and Smith, Digory and Woodhead, Simon and Ornelas, Nancy and Lan, Andrew},
  booktitle={Findings of NAACL 2024},
  pages={3067--3082},
  year={2024}
}

@article{koedinger2004real,
  title={The Real Story behind Story Problems: Effects of Representations on Quantitative Reasoning},
  author={Koedinger, Kenneth R. and Nathan, Mitchell J.},
  journal={The Journal of the Learning Sciences},
  volume={13},
  number={2},
  pages={129--164},
  year={2004}
}

@article{wang2020diagnostic,
  title={Diagnostic Questions: The NeurIPS 2020 Education Challenge},
  author={Wang, Zichao and Lamb, Angus and Saveliev, Evgeny and Cameron, Pashmina and Zaykov, Yordan and Hern{\'a}ndez-Lobato, Jos{\'e} Miguel and Turner, Richard E and Baraniuk, Richard G and Barton, Craig and Jones, Simon Peyton and Woodhead, Simon and Zhang, Cheng},
  journal={arXiv preprint arXiv:2007.12061},
  year={2020}
}

@book{rasch1960probabilistic,
  title={Probabilistic Models for Some Intelligence and Attainment Tests},
  author={Rasch, Georg},
  year={1960},
  publisher={University of Chicago Press}
}

@misc{yuan2026validstudentsimulationlarge,
  title={Towards Valid Student Simulation with Large Language Models},
  author={Yuan, Zhihao and Xiao, Yunze and Li, Ming and Xuan, Weihao and Tong, Richard and Diab, Mona and Mitchell, Tom},
  year={2026},
  note={arXiv:2601.05473}
}

@inproceedings{yaneva2024findings,
  title={Findings from the First Shared Task on Automated Prediction of Difficulty and Response Time for Multiple-Choice Questions},
  author={Yaneva, Victoria and North, Kai and Baldwin, Peter and Ha, Le An and Rezayi, Saed and Zhou, Yiyun and Ray Choudhury, Sagnik and Harik, Polina and Clauser, Brian},
  booktitle={Proceedings of BEA Workshop at NAACL 2024},
  pages={470--482},
  year={2024}
}

@inproceedings{hu2024quantifying,
  title={Quantifying the Persona Effect in {LLM} Simulations},
  author={Hu, Tiancheng and Collier, Nigel},
  booktitle={Proceedings of ACL 2024},
  pages={10289--10307},
  year={2024}
}

@inproceedings{scarlatos2023turning,
  title={Turning Large Language Models into Cognitive Models},
  author={Binz, Marcel and Schulz, Eric},
  booktitle={The Twelfth International Conference on Learning Representations},
  year={2024}
}

@inproceedings{haetal2019predicting,
  title={Predicting the Difficulty of Multiple Choice Questions in a High-stakes Medical Exam},
  author={Ha, Le An and Yaneva, Victoria and Baldwin, Peter and Mee, Janet},
  booktitle={Proceedings of BEA Workshop 2019},
  pages={11--20},
  year={2019}
}

@article{frances2014item,
  title={Item Response Theory for Measurement Validity},
  author={Yang, Frances M. and Kao, Solon T.},
  journal={Shanghai Archives of Psychiatry},
  volume={26},
  number={3},
  pages={171},
  year={2014}
}

@book{de2013theory,
  title={The Theory and Practice of Item Response Theory},
  author={De Ayala, Rafael Jaime},
  year={2013},
  publisher={Guilford Publications}
}

@article{butler2018multiple,
  title={Multiple-choice Testing in Education: Are the Best Practices for Assessment Also Good for Learning?},
  author={Butler, Andrew C},
  journal={Journal of Applied Research in Memory and Cognition},
  volume={7},
  number={3},
  pages={323--331},
  year={2018}
}

@inproceedings{lalor2019emnlp,
  title={Learning Latent Parameters without Human Response Patterns: Item Response Theory with Artificial Crowds},
  author={Lalor, John P and Wu, Hao and Yu, Hong},
  booktitle={Proceedings of EMNLP 2019},
  year={2019}
}

@article{hoerl1970ridge,
  title={Ridge Regression: Biased Estimation for Nonorthogonal Problems},
  author={Hoerl, Arthur E and Kennard, Robert W},
  journal={Technometrics},
  volume={12},
  number={1},
  pages={55--67},
  year={1970}
}

@article{beltagy2020longformer,
  title={Longformer: The Long-Document Transformer},
  author={Beltagy, Iz and Peters, Matthew E and Cohan, Arman},
  journal={arXiv preprint arXiv:2004.05150},
  year={2020}
}

@article{hurst2024gpt,
  title={GPT-4o System Card},
  author={Hurst, Aaron and Lerer, Adam and Goucher, Adam P and others},
  journal={arXiv preprint arXiv:2410.21276},
  year={2024}
}

@misc{anthropic2025opus45,
  title={Introducing Claude Opus 4.5},
  author={Anthropic},
  year={2025}
}

@misc{anthropic2025sonnet37,
  title={Claude 3.7 Sonnet},
  author={Anthropic},
  year={2025}
}
}

\end{document}